\documentclass[11pt,a4paper]{article}
\pdfoutput=1
\usepackage[english]{babel}
\usepackage[utf8]{inputenc}
\usepackage{jheppub}
\usepackage{amsmath}
\usepackage{nccmath}
\usepackage{mathtools}
\usepackage{amssymb}
\usepackage{amsfonts}
\usepackage{nameref}
\usepackage{latexsym}
\numberwithin{equation}{section}
\allowdisplaybreaks

\title{On a Gravity Dual to Flavored Topological Quantum Mechanics}
\author{Andrey Feldman}
\affiliation{Department of Particle Physics and Astrophysics, \\
Weizmann Institute of Science, Rehovot 7610001, Israel}
\emailAdd{andrey.feldman@weizmann.ac.il}
\abstract{In this paper, we propose a generalization of the $\mathrm{AdS_2/CFT_1}$ correspondence constructed by Mezei, Pufu and Wang in \cite{MezeiPufuWang}, which is the duality between 2d Yang-Mills theory with higher derivatives in the $\mathrm{AdS}_2$ background, and 1d topological quantum mechanics of two adjoint and two fundamental $\mathrm{U}(N)$ fields, governing certain protected sector of operators in 3d ABJM theory at the Chern-Simons level $k = 1$. We construct a holographic dual to a flavored generalization of the 1d quantum mechanics considered in \cite{MezeiPufuWang}, which arises as the effective field theory living on the intersection of stacks of $N$ D2-branes and $k$ D6-branes in the $\Omega$-background in Type IIA string theory, and describes the dynamics of the protected sector of operators in $\mathcal{N} = 4$ theory with $k$ fundamental hypermultiplets, having a holographic description as M-theory in the $\mathrm{AdS}_4 \times \mathrm{S}^7/\mathbb{Z}_k$ background. We compute the structure constants of the bulk theory gauge group, and construct a map between the observables of the boundary theory and the fields of the bulk theory.}

\begin{document}


\maketitle

\flushbottom

\selectlanguage{english}

\section{Introduction}

A holographic duality we consider in this paper is a generalization of the duality between the 1d topological matrix quantum mechanics (QM), which can be viewed as a subsector of the 3d $\mathcal{N} = 4$ gauge theory \cite{ChesterLeePufuYacoby,DedushenkoPufuYacoby,BeemPeelaersRastelli}, and a non-linear higher-derivative generalization of the Yang-Mills theory in the $\mathrm{AdS}_2$ background. The first example of such a duality was given in \cite{MezeiPufuWang}, where an equivalence of the 1d topological QM of interacting $\mathrm{U} (N)$ gauge field, a fundamental $Q$, anti-fundamental $\widetilde{Q}$, and two adjoint $X, \widetilde{X}$ scalars, and a non-linear Yang-Mills theory on $\mathrm{AdS}_2$ with the gauge group $\mathrm{SDiff \left( S^2 \right)}$ of area-preserving diffeomorphisms of the 2-sphere, was considered.

The QM theory we work with is the generalization of the theory studied in \cite{MezeiPufuWang} to the case of $k$ fundamental scalars, and non-vanishing Fayet–Iliopoulos (FI) term. This theory is particularly interesting, because it arises as the effective field theory living on the intersection of stacks of $N$ D2-branes and $k$ D6-branes in the $\Omega$-background in Type IIA string theory, with the gauge coupling $\Delta$ and the FI parameter $\epsilon$ identified with two equivariant parameters of the $\Omega$-deformation \cite{TwistedSupergravityAndItsQuantization,CostelloMTheoryOmegaBackground,HolographyAndKoszulDuality,GaiottoAbajian}.\footnote{This brane configuration is equivalent to a stack of M2-branes in the Taub-NUT background in the $\Omega$-deformed M-theory.}

The AdS/CFT correspondence \cite{MaldacenaHol,GKPHol,WittenHol} gives a dictionary between the single-trace conserved currents of the boundary theory and the gauge fields in the bulk higher-dimensional theory \cite{WittenHol}. For the case of one pair of fundamental/anti-fundamental scalars considered in \cite{MezeiPufuWang}, all operators involving $Q$'s can be expressed in terms of $X$'s and $\widetilde{X}$'s, and all the single-trace conserved currents are of the form
\begin{equation}
j^{p_1, \cdots, p_n}_{q_1, \cdots, q_n} = \mathrm{Tr} \left( X^{p_1} \widetilde{X}^{q_1} \cdots X^{p_n} \widetilde{X}^{q_n} \right),
\end{equation}
where $p_i$ and $q_j$ are some non-negative integer numbers. One can extract the structure constants of the bulk theory gauge group from the boundary 3-point function, as we will briefly review. For the case $k \neq 1$, there are the single-trace conserved currents which are non-scalars (i.e. they are traceless) under the $\mathrm{SU} (k)$, having the form
\begin{equation}
j^{p_1, \cdots, p_n,a}_{q_1, \cdots, q_n,b} =  \widetilde{Q}^a X^{p_1} \widetilde{X}^{q_1} \cdots X^{p_n} \widetilde{X}^{q_n} Q_b,
\end{equation}
and it implies that one cannot express them in terms of $X$'s. We consider the correlation functions of these operators in the 1d theory with $k \neq 1$, $\epsilon \neq 0$ in order to deduce the gauge group of the bulk theory.

\section{\boldmath $\mathrm{AdS}_2$ side}

Here we give a very brief introduction to the higher-derivative generalization of the Yang-Mills theory, closely following \cite{MezeiPufuWang}.

The gauge field dynamics in two dimensions is topological, so the gauge theory in the $\mathrm{AdS}_2$ background may be holographically dual to a 1d topological theory on the $\partial \mathrm{AdS}_2$.\footnote{More precisely, the bulk theory is quasi-topological, i.e. an arbitrary diffeomorphism on the boundary can be extended to an area-preserving diffeomorphism in the bulk.} We introduce the coordinates $\varphi$ and $r$, in which the 2d metric is of the form $\mathrm{d} s^2 = \mathrm{d} r^2 + \sinh^2 r \mathrm{d} \varphi^2$ (we put the $\mathrm{AdS}_2$ radius here to 1). The action governing the dynamics of the gauge field $A_{\mu} = \left( A_{\varphi}, A_r \right)$ of the bulk theory is
\begin{align} \label{eq:BulkAction}
\begin{split}
I_{\mathrm{2d}} = &\int \mathrm{d}^2 x \sqrt{g} \ \mathrm{Tr} \left[ \sum\limits_{n=2}^{\infty} \frac{d^{a_1 \cdots a_n}_n}{n!} F^{a_1} \cdots F^{a_n} \right] \\ &+\int \mathrm{d} \varphi \mathrm{Tr} \left[ \sum\limits_{n=2}^{\infty} \frac{1}{n!} \frac{A_{\varphi}^a d^{a a_2 \cdots a_n}_n}{n-1} \left( \frac{ A_{\varphi}^{a_2} \cdots A_{\varphi}^{a_n}}{\sinh^{n-1} R} - n F^{a_2} \cdots F^{a_n} \right) \right],
\end{split}
\end{align}
where $F^a = \frac{1}{2} \varepsilon^{\mu \nu} F^a_{\mu \nu}$, $R$ is the IR bulk cutoff, and $d^{a_1 \cdots a_n}_n$ are some totally symmetric invariant tensors of the bulk theory gauge group. In the original formulation of the duality given in \cite{MezeiPufuWang}, $d^{a_1 \cdots a_n}_n \sim 1/g_{YM}^2$. We will see that in our case the dependence of the bulk action on $g_{YM}$ is more general.

The solution of the bulk equations of motion can be written in the form
\begin{align}
\begin{split}
F_{r \varphi} &= U Q U^{-1} \sinh r, \\ A_{\varphi} &= U Q U^{-1} (\cosh r - 1) + \mathrm{i} U \partial_{\varphi} U^{-1}, \\ A_r &= \mathrm{i} U \partial_r U^{-1}.
\end{split}
\end{align}
Here $Q$ is an arbitrary constant Lie-algebra-valued matrix, and $U (r , \varphi)$ is an arbitrary Lie-group-valued matrix, which has a finite limit
$U(\infty, \varphi) \equiv u(\varphi)$ as $r \rightarrow \infty$, corresponding to large gauge transformations. The role of the non-normalizable mode, sourcing the conserved current operator $j^a \left( \varphi \right)$ in the boundary theory, is played by $a(\varphi) = A_{\varphi} + \lim\limits_{r \rightarrow \infty} F_{\varphi r}$. For the solution we gave, it has the form
\begin{equation}
a(\varphi) = \mathrm{i} u \partial_{\varphi} u^{-1} - u Q u^{-1}.
\end{equation}
Using this expression, one can write the action in terms of the source, which is useful to compute the correlation functions of currents, which can be compared with the boundary theory computations. The action $I_2^{\mathrm{on-shell}}[a]$ is
\begin{equation}
I_{\mathrm{2d}}^{\mathrm{on-shell}}[a] = - \int \mathrm{d}^2 x \sqrt{g} \ \mathrm{Tr} \left[\sum\limits_{n=2}^{\infty} \frac{d^{a_1 \cdots a_n}_n}{n!} Q^{a_1} \cdots Q^{a_n} \right],
\end{equation}
where $Q$ must be represented in terms of $a (\varphi)$ as
\begin{equation}
Q = \mathrm{i} u_0^{-1} \log \left( \hat{\mathrm{P}} \mathrm{exp} \left[ \mathrm{i} \int\limits_{- \pi}^{\pi} \mathrm{d} \varphi \ a(\varphi) \right] \right) u_0.
\end{equation}
The constant matrix $u_0$ is arbitrary, and cancels when we substitute $Q$ in the action, and $\hat{\mathrm{P}}$ is the path-ordering operator. The correlation functions of the boundary currents can then be computed as
\begin{equation} \label{eq:GeneralBulkCorrelator}
\left< j^{a_1}(\varphi_1) \cdots j^{a_n}(\varphi_n) \right> = (-1)^{n+1} \frac{\delta I_{2 \mathrm{d}}^{\mathrm{on-shell}}[a]}{\delta a_1(\varphi_1) \cdots \delta a_n(\varphi_n)}.
\end{equation}

\section{Field theory side}

\subsection{A one-dimensional topological quantum mechanics}

The field content of the 1d topological theory we are interested in is a $\mathrm{U} (N)$ gauge field $\mathcal{A}$, a $\mathrm{U} (N)$ fundamental field $Q$, an anti-fundamental field $\widetilde{Q}$, and a pair of adjoint fields $X$ and $\widetilde{X}$. The matter fields are anti-periodic when the theory is defined on a circle parameterized by $\varphi \in \left[- \pi, \pi \right)$. The dynamics is governed by the action
\begin{equation} \label{eq:FieldTheoryAction}
I_{\mathrm{1d}} = - \frac{1}{\Delta} \int \left\{\widetilde{Q}^{a} \left( \mathrm{d} + \mathcal{A} \right) Q_{a} + \mathrm{Tr} \left( \widetilde{X} \mathrm d X + \widetilde{X} [\mathcal{A}, X] \right) - \epsilon \mathrm{Tr} \mathcal{A} \right\}, 
\end{equation}
where $a$ runs from 1 to $k$, $\epsilon$ is the FI parameter, and $\Delta$ is the coupling constant. The fields $Q$ and $\widetilde{Q}$ transform under the flavor group $\mathrm{SU} (k)$ as $k$ and $\overline{k}$, respectively. The theory doesn't depend on the worldline metric, so it is topological, and it is natural to map the single-trace scalar gauge-invariant operators\footnote{We will see in what follows that the single-trace operators must be mixed with the double-trace ones in order to have a consistent bulk dual gauge theory.} to the gauge fields in the bulk, not to scalars, because the 2d scalar field theory is not topological. One can say that these operators must be considered as charges of conserved 1d currents.

The partition function of the theory is an obvious generalization of the result of \cite{MezeiPufuWang}, where the case of $\epsilon = 0$, and $k = 1$ is considered. It is given by
\begin{equation}
Z = \frac{1}{|\mathcal{W}|} \int\limits_{\text{Cartan of $\mathfrak{u} (N)$}} \mathrm{d} \sigma \ \text{det}'_{\text{adj}} \left( 2 \sinh (\pi \sigma) \right) Z_{\sigma},
\end{equation}
where $\mathcal{W}$ is the Weyl group of $\mathrm{U} (N)$, whose order is $|\mathcal{W}| = N!$, $\sigma$ parameterizes the Cartan subalgebra, and $\sigma = \text{diag} \left\{ \sigma_1, \cdots, \sigma_N \right\}$, $\text{det}'_{\text{adj}} \left( 2 \sinh (\pi \sigma) \right) = \prod\limits_{i<j} \left[ 2 \sinh \left( \pi \left( \sigma_i - \sigma_j \right) \right) \right]^2$, where the prime means that the zero modes are omitted, and $Z_{\sigma}$ is
\begin{equation}
Z_{\sigma} = \int \mathrm{D} Q \ \mathrm{D} \widetilde{Q} \ \mathrm{D} X \ \mathrm{D} \widetilde{X} \ \text{exp} \left[ \frac{1}{\Delta} \int \mathrm{d} \varphi \left\{\widetilde{Q}^{a} D_{\varphi}^F Q_{a} + \mathrm{Tr} \left( \widetilde{X} D_{\varphi}^A X \right) - \epsilon \mathrm{Tr} \sigma \right\} \right].
\end{equation}
Here the gauge choice $\mathcal{A}_{\varphi} = \sigma$ has been made, and the covariant derivatives in the fundamental $D_{\varphi}^F$ and adjoint $D_{\varphi}^A$ representations are defined as
\begin{equation}
D_{\varphi}^F Q = \partial_{\varphi} Q + \sigma Q, \qquad D_{\varphi}^A X = \partial_{\varphi} X + [\sigma, X].
\end{equation} 
The path integral $Z_{\sigma}$ is quadratic in the matter fields, and can be easily computed using the $\zeta$-function regularization. When it is done, the partition function takes the form
\begin{equation} \label{eq:matrixZ}
Z = \frac{1}{N! \ 2^{N}} \int \prod\limits_{i=1}^N \mathrm{d} \sigma_i \frac{\prod\limits_{i<j} \sinh^2 \left( \pi \left( \sigma_i - \sigma_j \right) \right)}{\prod\limits_{i,j=1}^N \cosh \left( \pi \left( \sigma_i - \sigma_j \right) \right) \prod\limits_{i=1}^N \left[ 2 \cosh \left( \pi \sigma_i \right) \right]^k} \mathrm{e}^{ - 2 \pi \frac{\epsilon}{\Delta} \sum\limits_{i=1}^N \sigma_i}.
\end{equation}

Introducing the parameter $\alpha$ such that $\epsilon = \frac{\alpha}{2} \Delta$, we re-write the integral in (\ref{eq:matrixZ}) as
\begin{equation} \label{eq:matrixZexponentiated}
\int\prod\limits_{i=1}^N \mathrm{d} \sigma_i \ \mathrm{e}^{-I_{eff}} = \int \prod\limits_{i=1}^N \mathrm{d} \sigma_i \ \mathrm{e}^{ - \sum\limits_{i<j} \log \left\{\coth^2 \left( \pi \left( \sigma_i - \sigma_j \right) \right) \right\} - k \sum\limits_{i=1}^N \log \left\{2 \cosh \left( \pi \sigma_i \right) \right\} - \alpha \pi \sum\limits_{i=1}^N \sigma_i}.
\end{equation}

\subsection{Large N computation}

\subsubsection{Partition function}

The matrix integral (\ref{eq:matrixZexponentiated}) and its generalizations with insertions, can be computed in the large $N$ approximation. We start our computation with the partition function.

If we make the change of variables $\sigma_i = \sqrt{N} x_i$, the spectrum of eigenvalues becomes dense in the variables $x_i$, and we can introduce the density of eigenvalues $\rho(x)$. After the replacement (see \cite{ChernSimonsReview} for a review)
\begin{equation}
\sum\limits_{i=1}^N V(\sigma_i) \rightarrow N \int \mathrm{d} x \ \rho(x) V (x),
\end{equation}
we get the effective action
\begin{align} \label{eq:SLargeN}
\begin{split}
I_{eff} = &\frac{N^2}{2} \int \mathrm{d} x \ \mathrm{d} y \ \rho(x) \rho(y) \log \left\{\coth^2 \left( \pi \sqrt{N} \left( x - y \right) \right) \right\} \\ &+ N \int \mathrm{d} x \ \rho(x) \left[ k  \log \left\{2 \cosh \left( \pi \sqrt{N} x \right) \right\} + \alpha \pi \sqrt{N} x \right],
\end{split}
\end{align}
which in the large $N$ limit can be written as
\begin{equation} \label{eq:SLargeNApprox}
I_{eff} = \pi N^{3/2} \int \mathrm{d} x \left[ \frac{1}{4} \rho(x)^2 + \left( k |x| + \alpha x \right) \rho(x) \right],
\end{equation}
using the large $N$ approximations
\begin{equation}
\log \left\{\coth^2 \left( \pi \sqrt{N} \left( x - y \right) \right) \right\} \approx \frac{\pi}{2 \sqrt{N}} \delta (x-y), \quad \log \left\{2 \cosh \left( \pi \sqrt{N} x \right) \right\} \approx \pi \sqrt{N} |x|.
\end{equation}

We extremize (\ref{eq:SLargeNApprox}) with respect to $\rho(x)$, imposing the additional constraint $\int \mathrm{d} x \ \rho(x) = 1$. The solution is
\begin{align}
\begin{split}
\rho(x)=\begin{cases}
\sqrt{2} \sqrt{\frac{k^2 - \alpha^2}{k}} - 2 \left( \alpha x + k |x| \right) + \mathcal{O} \left( \frac{1}{\sqrt{N}} \right), \qquad &x \in \left( -  \frac{1}{\sqrt{2 k}} \sqrt{\frac{k + \alpha}{k - \alpha}}, \frac{1}{\sqrt{2 k}} \sqrt{\frac{k - \alpha}{k + \alpha}} \right), \\
0, &\text{otherwise.}
\end{cases}
\end{split}
\end{align}
Here and in what follows, we put $| k | \geq | \alpha | $. If this inequality is not obeyed, it is hard to make sense of the model, because naively the path integral diverges.

Substituting $\rho(x)$ into (\ref{eq:SLargeNApprox}), we obtain\footnote{The theory with $\alpha = 0$ was also considered in \cite{MezeiPufu,GrassiMarino}.}
\begin{equation}
I_{eff} = \frac{\sqrt{2} \pi}{3} \sqrt{\frac{k^2 - \alpha^2}{k}} N^{\frac{3}{2}}.
\end{equation}

\subsubsection{Correlation functions of \boldmath $X$'s}

Typically, the holographic correspondence states that only the single-trace boundary operators have the bulk gauge fields as their duals, and the exchange of multi-trace operators in $n$-point functions should be automatically taken into account by the bulk dynamics. As we already mentioned, and will see in what follows, for the theories we consider this is almost the case, but the single-trace operators must be mixed with the double-trace operators.

In order to compute correlation functions of currents involving the fields $X$ and $\widetilde{X}$, it is convenient to introduce the notations $(X_1, X_2) = (\widetilde{X},X)$, and to rewrite the $X,\widetilde{X}$-dependent terms in (\ref{eq:FieldTheoryAction}) as
\begin{equation}
\int \mathrm{d} \varphi \mathrm{Tr} \left( \widetilde{X} D_{\varphi}^A X \right) = \frac{1}{2} \int \mathrm{d} \varphi \ \varepsilon^{IJ} \mathrm{Tr} \left( X_I \partial_{\varphi} X_J - \sigma \left[ X_I , X_J \right] \right),
\end{equation}
where $I,J = 1,2$. This allows us to write all the operators we are interested in as functions of
\begin{equation}
\mathbb{X} (\varphi,y) = y^1 \widetilde{X}(\varphi) + y^2 X(\varphi),
\end{equation}
where $\vec{y}$ is a 2d polarization vector. The basis of operators consists of all products of the form $\mathrm{Tr} \left[ \mathbb{X} (\varphi,y)^n \right]$
modulo trace relations that can be ignored in the large $N$ limit. We will focus on the single trace operators with even powers of $\mathbb{X}$, corresponding to integer spin under the $\mathfrak{su}(2)$, under which $(\widetilde{X},X)$ transform as a doublet. For convenience, we introduce the notation
\begin{equation}
\mathfrak{X}_{\ell} (\varphi,y) = \frac{1}{N^{\frac{\ell - 1}{2}}} \mathrm{Tr} \left[ \mathbb{X} (\varphi,y)^{2 \ell} \right].
\end{equation}
The operators $\mathfrak{X}_{\ell}$ contain only symmetric combinations of products of $X$ and $\widetilde{X}$. One can argue that it is enough to take only these combinations into account. Let's consider an operator with one anti-symmetrization $\mathrm{Tr} \left( \mathbb{X}^{2 \ell} [X, \widetilde{X}] \right)$. This operator vanishes. If we take the component with the highest spin projection in the expansion of $\mathrm{Tr} \left( \mathbb{X}^{2 \ell} [X, \widetilde{X}] \right)$, we get $\mathrm{Tr} \left( X^{2 \ell} [X, \widetilde{X}] \right)$, which is obviously zero by cyclicity of trace. The other components can be obtained by an $\mathfrak{su} (2)$ transformation, and thus  also vanish. The simplest case of an operator with more than one anti-symmetrization is $\mathrm{Tr} \left( [ X , \widetilde{X}]^2 \right)$. Using the $\sigma$ equations of motion, $Q_i \widetilde{Q}^j \sim [ X , \widetilde{X}]_i^{\ j}$, we get $\mathrm{Tr} \left( [ X , \widetilde{X}]^2 \right) \sim \widetilde{Q}^i [X, {\widetilde X}]_i^{\ j} Q_j \sim Q_i Q_j {\widetilde Q}^i {\widetilde Q}^j \sim \left\{ \mathrm{Tr} \left([X, {\widetilde X}] \right) \right\}^2 = 0$. We didn't work out the details of vanishing (or expressibility in terms of other operators) of more general operators, but validity of our assumption is confirmed by the fact that the Jacobi identities for the structure constants we extract from the 3-point functions, are satisfied, as we will see.

The $\mathfrak{su} (2)$ symmetry fixes the 2- and 3-point functions to be
\begin{align} \label{eq:Xcorrelators}
\begin{split}
\left< \mathfrak{X}_{\ell_1} (\varphi_1 , y_1) \mathfrak{X}_{\ell_2} (\varphi_2, y_2) \right> = & \ B_{\ell_1}^{\mathfrak{XX}} \delta_{\ell_1 \ell_2} \left< y_1, y_2 \right>^{2 \ell_1}, \\
\left< \mathfrak{X}_{\ell_1} (\varphi_1 , y_1) \mathfrak{X}_{\ell_2} (\varphi_2, y_2) \mathfrak{X}_{\ell_3} (\varphi_3, y_3) \right> = & \ C_{\ell_1 \ell_2 \ell_3}^{\mathfrak{XXX}} \left< y_1, y_2 \right>^{L_{12,3}} \left< y_1, y_3 \right>^{L_{13,2}} \left< y_2, y_3 \right>^{L_{23,1}} \\ &\times \left( \mathrm{sgn} \varphi_{21} \right)^{L_{12,3}} \left( \mathrm{sgn} \varphi_{31} \right)^{L_{13,2}} \left( \mathrm{sgn} \varphi_{32} \right)^{L_{23,1}},
\end{split}
\end{align}
where $\varphi_{ij} = \varphi_i - \varphi_j$, $\left< y_1, y_2 \right> = \varepsilon_{IJ} y_1^I y_2^J$ with $\varepsilon_{12} = - 1$, and $L_{ij,k} = \ell_i + \ell_j - \ell_k$.

To compute the correlation functions, we also need the propagator for the field $\mathbb{X}$ at fixed $\sigma$. The propagator for a scalar field $\Phi$ in the representation $\mathcal{R}$ is given by
\begin{equation}
\left< \Phi (\varphi_1) \widetilde{\Phi} (\varphi_2) \right>_{\sigma} = - \frac{\Delta}{2} \left( \tanh (\pi \sigma) + \mathrm{sgn} (\varphi_{12}) \textbf{1} \right) \mathrm{e}^{- \sigma \varphi_{12}},
\end{equation}
where $\textbf{1}$ is the $\mathrm{dim} \mathcal{R} \times \mathrm{dim} \mathcal{R}$ unit matrix, and $\sigma$ is also must be thought of as the $\mathrm{dim} \mathcal{R} \times \mathrm{dim} \mathcal{R}$ matrix. When the fields sit at the same point, the propagator is defined as
\begin{equation}
\left< \Phi (\varphi) \widetilde{\Phi} (\varphi) \right>_{\sigma} = - \frac{\Delta}{2} \tanh (\pi \sigma).
\end{equation}
It follows that for the adjoint field $\mathbb{X}$ one gets
\begin{equation}
\left< \mathbb{X}_i^{\ j} (\varphi_1, y_1) \mathbb{X}_k^{\ l} (\varphi_2 , y_2) \right>_{\sigma} = \left< y_1 , y_2 \right> \delta_i^l \delta_k^j G_{\sigma_{ij}}(\varphi_{12}),
\end{equation}
where $\sigma_{ij} = \sigma_i - \sigma_j$, and (see \cite{DedushenkoPufuYacoby})
\begin{equation} \label{eq:UniversalPropagator}
G_{\sigma} (\varphi) = - \frac{\Delta}{2} \left( \tanh (\pi \sigma) + \mathrm{sgn} (\varphi) \right) \mathrm{e}^{- \sigma \varphi} =  - \frac{\Delta}{2} \mathrm{sgn}(\varphi) \frac{\mathrm{e}^{\mathrm{sgn} \left( \varphi \right) \pi \sigma}}{\cosh (\pi \sigma)} \mathrm{e}^{- \sigma \varphi}.
\end{equation}

It is easy to compute the 2-point function at arbitrary $\ell$, generalizing the computation of \cite{MezeiPufuWang}. The result for the coefficient $B_{\ell}^{\mathfrak{XX}}$ is
\begin{align}
\begin{split}
B_{\ell}^{\mathfrak{XX}} &= 2 \ell N^{\ell + 1} \left( \frac{\Delta}{2} \right)^{2 \ell} \int \left[ \prod\limits_{a = 1}^{2 \ell} \mathrm{d} x_a \ \rho \left(x_a \right) \frac{1}{\cosh \left( \pi \sqrt{N} \left( x_a - x_{a+1} \right) \right)} \right] \\ &= \frac{2^{\ell + \frac{1}{2}}}{\sqrt{\pi}} \left( \frac{\Delta}{2} \right)^{2 \ell} \frac{\Gamma ( \ell + 1)}{\Gamma \left( \ell + \frac{3}{2} \right)} \left( \frac{k^2 - \alpha^2}{k} \right)^{\ell - \frac{1}{2}} N^{\frac{3}{2}} = \frac{2}{\sqrt{\pi}} \left( \frac{\Delta}{2} \right)^{2 \ell} \frac{\Gamma ( \ell + 1)}{\Gamma \left( \ell + \frac{3}{2} \right)} \rho(0)^{2 \ell - 1} N^{\frac{3}{2}},
\end{split}
\end{align}
where we defined $x_{2 \ell + 1} \equiv x_1$, and used the large $N$ approximation
\begin{equation}
\prod\limits_{a =1}^{2 \ell} \frac{1}{ \cosh \left( \pi \sqrt{N} \left( x_a - x_{a+1} \right) \right)} \approx \frac{1}{\sqrt{\pi}} \frac{\Gamma (\ell)}{\Gamma \left( \ell + \frac{1}{2} \right)} \frac{1}{N^{\ell - \frac{1}{2}}} \prod\limits_{a=1}^{2 \ell - 1} \delta \left( x_a - x_{a+1} \right).
\end{equation}
Here the coefficient on the right hand side can be obtained by integrating both sides with respect to $2 \ell - 1$ $x$'s, using the Fourier transform of $\frac{1}{\cosh \left( x_a - x_{a+1} \right)}$.

Generalizing then the result of \cite{MezeiPufuWang} for the 3-point function, one gets for the coefficient $C_{ \ell_1 \ell_2 \ell_3}^{\mathfrak{XXX}}$ the expression
\begin{align} \label{eq:XXXIntegral}
\begin{split}
&C_{ \ell_1 \ell_2 \ell_3}^{\mathfrak{XXX}} = \frac{8 \ell_1 \ell_2 \ell_3 N^{\ell_1 + \ell_2 + \ell_3 -1}}{N^{\frac{\ell_1 + \ell_2 + \ell_3-3}{2}}} \left( \frac{\Delta}{2} \right)^{\ell_1 + \ell_2 + \ell_3} \int \mathrm{d} x \ \mathrm{d} y \ \cosh \left(\pi \sqrt{N} \left(x - y \right) \right) \rho (x) \rho(y) \\ &\times \frac{\prod\limits_{i=1}^{L_{12,3}-1} \mathrm{d} u_i \ \rho \left( u_i \right)}{\cosh \left( \pi \sqrt{N} \left( x - u_1 \right) \right) \left[ \prod\limits_{i=1}^{L_{12,3}-2} \cosh \left( \pi \sqrt{N} \left( u_i - u_{i+1} \right) \right) \right] \cosh \left( \pi \sqrt{N} \left( u_{L_{12,3}-1} - y \right) \right)} \\ &\times \frac{\prod\limits_{j=1}^{L_{13,2}-1} \mathrm{d} v_j \ \rho \left( v_j \right)}{\cosh \left( \pi \sqrt{N} \left( x - v_1 \right) \right) \left[ \prod\limits_{j=1}^{L_{13,2}-2} \cosh \left( \pi \sqrt{N} \left( v_i - v_{j+1} \right) \right) \right] \cosh \left( \pi \sqrt{N} \left( v_{L_{13,2}-1} - y \right) \right)} \\ &\times \frac{\prod\limits_{k=1}^{L_{23,1}-1} \mathrm{d} w_k \ \rho \left( w_k \right)}{\cosh \left( \pi \sqrt{N} \left( x - w_1 \right) \right) \left[ \prod\limits_{k=1}^{L_{23,1}-2} \cosh \left( \pi \sqrt{N} \left( w_k - w_{k+1} \right) \right) \right] \cosh \left( \pi \sqrt{N} \left( w_{L_{23,1}-1} - y \right) \right)}.
\end{split}
\end{align}

Evaluating the integral in the large $N$ limit, we obtain
\begin{align}
\begin{split}
C_{ \ell_1 \ell_2 \ell_3}^{\mathfrak{XXX}} = 2 \frac{\rho(0)^{\ell_1 + \ell_2 + \ell_3 - 2}}{\sqrt{\pi}} \left( \frac{\Delta}{2} \right)^{\ell_1 + \ell_2 + \ell_3} N^{\frac{3}{2}} \left[ \frac{8 \ell_1 \ell_2 \ell_3}{\ell_1 + \ell_2 + \ell_3}  \mathcal{I} \left( \ell_1 , \ell_2, \ell_3 \right) \right],
\end{split}
\end{align}
where
\begin{equation}
\mathcal{I} \left( \ell_1 , \ell_2, \ell_3 \right) = \frac{\Gamma \left( \ell_1 \right) \Gamma \left( \ell_2 \right) \Gamma \left( \ell_3 \right)}{\Gamma \left( \frac{\ell_1 + \ell_2 - \ell_3}{2} + \frac{1}{2} \right) \Gamma \left( \frac{\ell_1 + \ell_3 - \ell_2}{2} + \frac{1}{2} \right) \Gamma \left( \frac{\ell_2 + \ell_3 - \ell_1}{2} + \frac{1}{2} \right) \Gamma \left( \frac{\ell_1 + \ell_2 + \ell_3}{2} \right)}.
\end{equation}
Here $\ell_1$, $\ell_2$, and $\ell_3$ must obey the triangle inequality $|\ell_1 - \ell_2| \leq \ell_3 \leq \ell_1 + \ell_2$; otherwise, the constant $C_{ \ell_1 \ell_2 \ell_3}^{\mathfrak{XXX}}$ vanishes.

It is easy to see that if we re-scale $\mathfrak{X}_{\ell} \rightarrow \hat{\mathfrak{X}}_{\ell}$ in such a way that
\begin{equation}
\left< \hat{\mathfrak{X}}_{\ell_1} \left( \varphi_1 , y_1 \right) \hat{\mathfrak{X}}_{\ell_2} \left( \varphi_2 , y_2 \right) \right> = \delta_{\ell_1 \ell_2},
\end{equation}
the 3-point function of the re-scaled operators is proportional to $ \left( \frac{k}{k^2 - \alpha^2} \right)^{\frac{1}{4}} N^{ - \frac{3}{4}}$, and it is natural to introduce the bulk gauge coupling
\begin{equation}
\frac{1}{g^2_{\mathrm{YM}}} = \sqrt{\frac{k^2 - \alpha^2}{k}} N^{\frac{3}{2}}.
\end{equation}
The correlation functions of the original operators scale as
\begin{align}
\left< \mathfrak{X}_{\ell_1} (\varphi_1 , y_1) \mathfrak{X}_{\ell_2} (\varphi_2, y_2) \right> &\sim 1/g_{YM}^2, \\ \left< \mathfrak{X}_{\ell_1} (\varphi_1 , y_1) \mathfrak{X}_{\ell_2} (\varphi_2, y_2) \mathfrak{X}_{\ell_3} (\varphi_3, y_3)  \right> &\sim 1/g_{YM}^2.
\end{align}

We didn't manage to compute the general connected 4-point function, but its $N$-dependence can be found, and it happens that it scales as
\begin{equation}
\left< \mathfrak{X}_{\ell_1} (\varphi_1 , y_1) \mathfrak{X}_{\ell_2} (\varphi_2, y_2) \mathfrak{X}_{\ell_3} (\varphi_3, y_3) \mathfrak{X}_{\ell_4} (\varphi_4, y_4) \right>_c \sim N^{\frac{3}{2}} \sim 1/g_{YM}^2.
\end{equation}
It confirms our holographic identification (\ref{eq:GeneralBulkCorrelator}) with $d^{a_1 \cdots a_n}_n \sim 1/g_{YM}^2$.

This dependence of the 2d bulk theory action and correlation functions on $N$ makes perfect sense. We expect that the fields $\mathfrak{X}_{\ell m}$ arise from the Kaluza-Klein modes of the 11d supergravity in the $\mathrm{AdS}_4 \times \mathrm{S}^7$ background, whose action behaves as $S_{11d} \sim \frac{1}{\ell_{Pl}^9} \int \mathrm{d}^{11} x \sqrt{g} R + \cdots$, and given the identification $\ell_{Pl} \sim N^{- \frac{1}{6}}$ in 3d/11d holographic duality \cite{ABJM}, the aforementioned $N^{\frac{3}{2}}$ behavior matches precisely the higher-dimensional one.

\subsubsection{Correlation functions involving \boldmath $Q$'s}

Since here we are interested in the single-trace currents only, we consider the following operators:
\begin{equation}
\mathfrak{Q}^A_{\ell} (\varphi,y) = \frac{\mathrm{i}}{N^{\frac{\ell}{2}}} \left( T^A \right)_a^{\ b} \widetilde{Q}^{a,i_1} \mathbb{X}_{i_1}^{\ i_2} \cdots \mathbb{X}_{i_{2 \ell}}^{\ i_{2 \ell + 1}} Q_{b, i_{2 \ell} + 1} = \frac{\mathrm{i}}{N^{\frac{\ell}{2}}} \left( T^A \right)_a^{\ b} \widetilde{Q}^a \mathbb{X}^{2 \ell} Q_b,
\end{equation}
where $T^A$ are the generators of $\mathrm{SU}(k)$, and the normalization is chosen in such a way that all correlators scale as the same power of $N$, which is necessary to have a gravity dual interpretation.

To compute the correlation functions of $\mathfrak{Q}_{\ell}$'s, we need the propagator of $Q$ and $\widetilde{Q}$, which is of the form
\begin{equation}
\left< Q (\varphi_1)_{a,i} \widetilde{Q} (\varphi_2)^{b,j} \right>_{\sigma} = - \frac{\Delta}{2} \delta_i^j \delta_a^b \mathrm{sgn}(\varphi_{12}) \frac{\mathrm{e}^{ \mathrm{sgn} \left( \varphi_{12} \right) \pi \sigma_i}}{\cosh (\pi \sigma_i)} \mathrm{e}^{- \sigma_i \varphi_{12}},
\end{equation}
It is not hard to obtain a general expression for the 2-point function, which is
\begin{align}
\begin{split}
&\left< \mathfrak{Q}^A_{\ell_1} (\varphi_1 , y_1) \mathfrak{Q}^B_{\ell_2} (\varphi_2,y_2) \right> = \delta_{\ell_1 \ell_2} \frac{1}{N^{\frac{\ell_1 + \ell_2}{2}}} \left( \frac{\Delta}{2} \right)^2 \mathrm{Tr} \left( T^A T^B \right) \\ &\times \sum\limits_{i,j} \frac{\mathrm{e}^{\mathrm{sgn} \left( \varphi_{12} \right) \pi \sigma_{ji}} \mathrm{e}^{ - \varphi_{12} \sigma_{ji}}}{\cosh \left( \pi \sigma_i \right) \cosh \left( \pi \sigma_j \right)} \left< \left[ \mathbb{X} (\varphi_1 , y_1)^{2 \ell_1} \right]_i^{\ j} \left[ \mathbb{X} (\varphi_2 , y_2)^{2 \ell_2} \right]_j^{\ i} \right>_{\sigma}.
\end{split}
\end{align}
After computing the sum in the large $N$ limit, one gets
\begin{equation} \label{eq:2pointQ}
\left< \mathfrak{Q}^A_{\ell} (\varphi_1 , y_1) \mathfrak{Q}^B_{\ell} (\varphi_2,y_2) \right> = \mathrm{Tr} \left( T^A T^B \right) \left( \frac{\Delta}{2} \right)^{2 \ell + 2} \rho(0)^{2 \ell + 1} \frac{1}{\sqrt{\pi}} \frac{\Gamma ( \ell + 1)}{\Gamma \left( \ell + \frac{3}{2} \right)} \sqrt{N} \left< y_1 , y_2 \right>^{2 \ell}.
\end{equation}
We can write (\ref{eq:2pointQ}) in the form similar to the first equation in (\ref{eq:Xcorrelators}), namely
\begin{equation} \label{eq:2pointnormalized}
\left< \mathfrak{Q}^A_{\ell_1} (\varphi_1 , y_1) \mathfrak{Q}^B_{\ell_2} (\varphi_2,y_2) \right> = \delta_{\ell_1 \ell_2}\mathrm{Tr} \left( T^A T^B \right) B_{\ell_1}^{\mathfrak{QQ}} \left< y_1 , y_2 \right>^{2 \ell_1},
\end{equation}
where
\begin{equation}
B_{\ell}^{\mathfrak{QQ}} = \frac{\rho(0)^2}{2} \left( \frac{\Delta}{2} \right)^2 \frac{1}{N} B_{\ell}^{\mathfrak{XX}},
\end{equation}
which means that the 2-point function of $\mathfrak{Q}_{\ell}$'s differs from the 2-point function of $\mathfrak{X}_{\ell}$'s only in an $\ell$-independent factor.

The 3-point function $\left< \mathfrak{Q}_{\ell_1}^A (\varphi_1 , y_1) \mathfrak{Q}_{\ell_2}^B (\varphi_2,y_2) \mathfrak{Q}_{\ell_3}^C (\varphi_3 , y_3) \right>$ can be written as
\begin{align} \label{eq:3pointfunction}
\begin{split}
\left< \mathfrak{Q}_{\ell_1}^A (\varphi_1 , y_1) \mathfrak{Q}_{\ell_2}^B (\varphi_2,y_2) \mathfrak{Q}_{\ell_3}^C (\varphi_3 , y_3) \right> = &\mathrm{i} \mathrm{Tr} \left( T^A \left[ T^B , T^C \right] \right) C_{\ell_1 \ell_2 \ell_3}^{\mathfrak{QQQ}} \\ &\times \left< y_1, y_2 \right>^{L_{12,3}} \left< y_1, y_3 \right>^{L_{13,2}} \left< y_2, y_3 \right>^{L_{23,1}} \\ &\times \left( \mathrm{sgn} \varphi_{21} \right)^{L_{12,3}+1} \left( \mathrm{sgn} \varphi_{31} \right)^{L_{13,2}+1} \left( \mathrm{sgn} \varphi_{32} \right)^{L_{23,1}+1},
\end{split}
\end{align}
where
\begin{align}
\begin{split}
&C_{\ell_1 \ell_2 \ell_3}^{\mathfrak{QQQ}} = \frac{N^{\ell_1 + \ell_2 + \ell_3 + 1}}{N^{\frac{\ell_1 + \ell_2 + \ell_3}{2}}} \left( \frac{\Delta}{2} \right)^{\ell_1 + \ell_2 + \ell_3+3} \int \mathrm{d} x \ \cosh \left(\pi \sqrt{N} x \right) \rho (x) \\ &\times \frac{\prod\limits_{i=1}^{L_{12,3}} \mathrm{d} u_i \ \rho \left( u_i \right)}{\cosh \left( \pi \sqrt{N} u_1 \right) \left[ \prod\limits_{i=1}^{L_{12,3}-1} \cosh \left( \pi \sqrt{N} \left( u_i - u_{i+1} \right) \right) \right] \cosh \left( \pi \sqrt{N} \left( u_{L_{12,3}} - x \right) \right)} \\ &\times \frac{ \prod\limits_{j=1}^{L_{13,2}} \mathrm{d} v_j \ \rho \left( v_j \right)}{\cosh \left( \pi \sqrt{N} v_1 \right) \left[ \prod\limits_{j=1}^{L_{13,2}-1} \cosh \left( \pi \sqrt{N} \left( v_j - v_{j+1} \right) \right) \right] \cosh \left( \pi \sqrt{N} \left( v_{L_{13,2}} - x \right) \right)} \\ &\times \frac{ \prod\limits_{k=1}^{L_{23,1}} \mathrm{d} w_k \ \rho \left( w_k \right)}{\cosh \left( \pi \sqrt{N} w_1 \right) \left[ \prod\limits_{k=1}^{L_{23,1}-1} \cosh \left( \pi \sqrt{N} \left( w_k - w_{k+1} \right) \right) \right] \cosh \left( \pi \sqrt{N} \left( w_{L_{23,1}} - x \right) \right)}.
\end{split}
\end{align}
Approximating the rapidly varying part of the integral (which doesn't include densities $\rho$) by
\begin{equation}
K^{\mathfrak{QQQ}} \delta(x) \left( \prod\limits_{i=1}^{L_{12,3}} \delta \left( u_i \right) \right) \left( \prod\limits_{j=1}^{L_{13,2}} \delta \left( v_j \right) \right) \left( \prod\limits_{k=1}^{L_{23,1}} \delta \left( w_k \right) \right),
\end{equation}
and integrating both sides with respect to $x$, $u_i$, $v_j$, and $w_k$ to find the constant $K$, we get
\begin{align} \label{eq:Ccoefficient}
\begin{split}
C_{\ell_1 \ell_2 \ell_3}^{\mathfrak{QQQ}} = \frac{\rho(0)^{ \ell_1 + \ell_ 2 + \ell_3 + 1}}{\sqrt{\pi}} \left( \frac{\Delta}{2} \right)^{\ell_1 + \ell_ 2 + \ell_3 + 3} \sqrt{N} \mathcal{I} \left( \ell_1 +1 , \ell_2 + 1 , \ell_3 + 1 \right).
\end{split}
\end{align}

The 3-point function $\left< \mathfrak{Q}_{\ell_1}^A (\varphi_1 , y_1) \mathfrak{Q}_{\ell_2}^B (\varphi_2,y_2) \mathfrak{X}_{\ell_3} (\varphi_3 , y_3) \right>$ has the form
\begin{align}
\begin{split}
\left< \mathfrak{Q}_{\ell_1}^A (\varphi_1 , y_1) \mathfrak{Q}_{\ell_2}^B (\varphi_2,y_2) \mathfrak{X}_{\ell_3} (\varphi_3 , y_3) \right> = &\mathrm{Tr} \left( T^A T^B \right) C_{\ell_1 \ell_2 \ell_3}^{\mathfrak{QQX}} \\ &\times \left< y_1, y_2 \right>^{L_{12,3}} \left< y_1, y_3 \right>^{L_{13,2}} \left< y_2, y_3 \right>^{L_{23,1}} \\ &\times \left( \mathrm{sgn} \varphi_{21} \right)^{L_{12,3}} \left( \mathrm{sgn} \varphi_{31} \right)^{L_{13,2}} \left( \mathrm{sgn} \varphi_{32} \right)^{L_{23,1}}.
\end{split}
\end{align}
Here $C_{\ell_1 \ell_2 \ell_3}^{\mathfrak{QQX}}$ is given by
\begin{align}
\begin{split}
&C_{\ell_1 \ell_2 \ell_3}^{\mathfrak{QQX}} = 2 \ell_3 \left( L_{12,3} + 1 \right) \frac{ N^{L}}{N^{\frac{L-1}{2}}} \left( \frac{\Delta}{2} \right)^{L+2} \int \frac{\mathrm{d} x \ \mathrm{d} y \ \mathrm{d} z \ \cosh \left(\pi \sqrt{N} \left(x - y \right) \right) \rho (x) \rho (y) \rho (z)}{\cosh \left( \pi \sqrt{N} x \right) \cosh \left( \pi \sqrt{N} z \right)} \\ &\times \frac{\prod\limits_{i=1}^{L_{12,3}-1} \mathrm{d} u_i \ \rho \left( u_i \right)}{\cosh \left( \pi \sqrt{N} \left( z - u_1 \right) \right) \left[ \prod\limits_{i=1}^{L_{12,3}-2} \cosh \left( \pi \sqrt{N} \left( u_i - u_{i+1} \right) \right) \right] \cosh \left( \pi \sqrt{N} \left( u_{L_{12,3}-1} - y \right) \right)} \\ &\times \frac{\prod\limits_{j=1}^{L_{13,2}-1} \mathrm{d} v_j \ \rho \left( v_j \right)}{\cosh \left( \pi \sqrt{N} \left( x - v_1 \right) \right) \left[ \prod\limits_{j=1}^{L_{13,2}-2} \cosh \left( \pi \sqrt{N} \left( v_i - v_{j+1} \right) \right) \right] \cosh \left( \pi \sqrt{N} \left( v_{L_{13,2}-1} - y \right) \right)} \\ &\times \frac{\prod\limits_{k=1}^{L_{23,1}-1} \mathrm{d} w_k \ \rho \left( w_k \right)}{\cosh \left( \pi \sqrt{N} \left( x - w_1 \right) \right) \left[ \prod\limits_{k=1}^{L_{23,1}-2} \cosh \left( \pi \sqrt{N} \left( w_k - w_{k+1} \right) \right) \right] \cosh \left( \pi \sqrt{N} \left( w_{L_{23,1}-1} - y \right) \right)},
\end{split}
\end{align}
where the factor $2 \ell_3 \left( L_{12,3} + 1 \right)$ counts different contractions which contribute at the leading order in $N$. Again, approximating the $\rho$-independent part of the integral by
\begin{equation}
K^{\mathfrak{QQX}} \delta(x) \delta(y) \delta(z) \left( \prod\limits_{i=1}^{L_{12,3}} \delta \left( u_i \right) \right) \left( \prod\limits_{j=1}^{L_{13,2}} \delta \left( v_j \right) \right) \left( \prod\limits_{k=1}^{L_{23,1}} \delta \left( w_k \right) \right),
\end{equation}
and integrating both sides with respect to $x$, $y$, $z$, $u_i$, $v_j$, and $w_k$ to find the constant $K^{\mathfrak{QQX}}$, we get
\begin{align} \label{eq:Dcoefficient}
\begin{split}
C_{\ell_1 \ell_2 \ell_3}^{\mathfrak{QQX}} &= \frac{\rho(0)^{\ell_1 + \ell_ 2 + \ell_3}}{\sqrt{\pi}} \left( \frac{\Delta}{2} \right)^{\ell_1 + \ell_ 2 + \ell_3 + 2} \sqrt{N} \left[ 2 \ell_3 \left( \ell_1 + \ell_2 - \ell_3 + 1 \right) \mathcal{I} \left( \ell_1 + 1 , \ell_2 + 1 , \ell_3 \right) \right] \\ &= \frac{\rho(0)^{2}}{2} \left( \frac{\Delta}{2} \right)^{2} \frac{1}{N} C_{\ell_1 \ell_2 \ell_3}^{\mathfrak{XXX}}.
\end{split}
\end{align}

This relation between $C_{\ell_1 \ell_2 \ell_3}^{\mathfrak{QQX}}$ and $C_{\ell_1 \ell_2 \ell_3}^{\mathfrak{XXX}}$ can be also obtained as follows. For a trivial flavor symmetry ($k = 1$) and $\alpha = 0$, the operators $\mathfrak{Q}_{\ell}$ are trivial in the chiral ring in the parent ABJM theory, so their correlators must vanish. In this case, contractions of $Q$ and $\widetilde{Q}$ within the same $\mathfrak{Q}_{\ell}$ also contribute to $C_{\ell_1 \ell_2 \ell_3}^{\mathfrak{QQX}}$, and must cancel the contribution of contractions between $Q$ at the different points, which were considered in our computation. Taking these self-contractions into account leads to multiplication of the integrand in (\ref{eq:XXXIntegral}) by $- \frac{1}{N} \left( \frac{\Delta}{2} \right)^2 \tanh \left( \pi \sigma_i \right) \tanh \left( \pi \sigma_m \right)$ for various $i$ and $m$. This factor can be written as
\begin{equation} \label{eq:tanh}
- \frac{1}{N} \left( \frac{\Delta}{2} \right)^2 \tanh \left( \pi \sigma_i \right) \tanh \left( \pi \sigma_m \right) = - \frac{1}{N} \left( \frac{\Delta}{2} \right)^2 \left[ 1 - \frac{\cosh \left( \pi \sigma_{im} \right)}{\cosh \left( \pi \sigma_i \right) \cosh \left( \pi \sigma_m \right)} \right],
\end{equation}
but the second term in the square brackets leads to additional $1/\sqrt{N}$ suppression, because of $\cosh$'s in the denominator, nailing all the $\sigma$'s to zero. One finally gets
\begin{equation}
C_{\ell_1 \ell_2 \ell_3}^{\mathfrak{QQX}} = - \left( \frac{\Delta}{2} \right)^2 \frac{1}{N} C_{\ell_1 \ell_2 \ell_3}^{\mathfrak{XXX}},
\end{equation}
which indeed cancels our (\ref{eq:Dcoefficient}) if $k=1$ and $\alpha = 0$.

A computation of the general connected 4-point function is again very cumbersome, but the $N$-dependence is found easily, and is of the form
\begin{align}
\left< \mathfrak{Q}_{\ell_1}^A (\varphi_1 , y_1) \mathfrak{Q}_{\ell_2}^B (\varphi_2,y_2) \mathfrak{Q}_{\ell_3}^C (\varphi_3 , y_3) \mathfrak{Q}_{\ell_4}^D (\varphi_4 , y_4) \right>_c &\sim \sqrt{N}, \\ \left< \mathfrak{Q}_{\ell_1}^A (\varphi_1 , y_1) \mathfrak{Q}_{\ell_2}^B (\varphi_2,y_2) \mathfrak{Q}_{\ell_3}^C (\varphi_3 , y_3) \mathfrak{X}_{\ell_4} (\varphi_4 , y_4) \right>_c &\sim \sqrt{N}, \\ \left< \mathfrak{Q}_{\ell_1}^A (\varphi_1 , y_1) \mathfrak{Q}_{\ell_2}^B (\varphi_2,y_2) \mathfrak{X}_{\ell_3} (\varphi_3 , y_3) \mathfrak{X}_{\ell_4} (\varphi_4 , y_4) \right>_c &\sim \sqrt{N}.
\end{align}

It suggests that in the bulk theory $d^{a_1 \cdots a_n}_n \sim 1/g_{YM}^2$ if all $a_n$ are $\mathfrak{X}$-indices, and $d^{a_1 \cdots a_n}_n \sim 1/g_{YM}^{\frac{2}{3}}$ if at least one of $a_n$ is an $\mathfrak{Q}$-index.

\subsubsection{A pure \boldmath $\mathfrak{Q}$ sector}

The correlation functions of the $\mathfrak{Q}_0^A$ sector are the following:
\begin{align}
\begin{split}
\left< \mathfrak{Q}^A_0 (\varphi_1) \mathfrak{Q}^B_0 (\varphi_2) \right> = &\mathrm{Tr} \left( T^A T^B \right) \left( \frac{\Delta}{2} \right)^2 \frac{2 \rho(0)}{\pi} \sqrt{N}, \\
\left< \mathfrak{Q}^A_0 (\varphi_1) \mathfrak{Q}^B_0 (\varphi_2) \mathfrak{Q}^C_0 (\varphi_3) \right> = &\mathrm{i} \mathrm{Tr} \left( T^A \left[ T^B , T^C \right] \right) \mathrm{sgn} \left( \varphi_{12} \varphi_{23} \varphi_{31} \right) \left( \frac{\Delta}{2} \right)^3 \frac{2 \rho(0)}{\pi} \sqrt{N}, \\
\left< \mathfrak{Q}_0^A (\varphi_1) \mathfrak{Q}_0^B (\varphi_2) \mathfrak{Q}_0^C (\varphi_3) \mathfrak{Q}_0^D (\varphi_4) \right> = & \mathrm{Tr} \left( T^A T^C T^D T^B + T^A T^B T^D T^C \right) \mathrm{sgn} \left( \varphi_{12} \varphi_{24} \varphi_{43} \varphi_{31} \right) \\ &\times \sum\limits_i \frac{\mathrm{e}^{\left( \mathrm{sgn} \varphi_{12} + \mathrm{sgn} \varphi_{24} + \mathrm{sgn} \varphi_{43} + \mathrm{sgn} \varphi_{31} \right) \pi \sigma_i}}{\cosh^4 \left( \pi \sigma_i \right)} \\ &+ \mathrm{permutations}.
\end{split}
\end{align}

For the simplest case of $G = \mathrm{SU} (2)$, one gets
\begin{align}
\begin{split}
&\left< \mathfrak{Q}^A_0 (\varphi_1) \mathfrak{Q}^B_0 (\varphi_2) \right> = \delta^{AB} \left( \frac{\Delta}{2} \right)^2 \frac{2 \rho(0)}{\pi} \sqrt{N}, \\
&\left< \mathfrak{Q}^A_0 (\varphi_1) \mathfrak{Q}^B_0 (\varphi_2) \mathfrak{Q}^C_0 (\varphi_3) \right> = - \varepsilon^{ABC} \left( \frac{\Delta}{2} \right)^3 \frac{2 \rho(0)}{\pi} \sqrt{N} \mathrm{sgn} \left( \varphi_{12} \varphi_{23} \varphi_{31} \right), \\
&\left< \mathfrak{Q}_0^A (\varphi_1) \mathfrak{Q}_0^B (\varphi_2) \mathfrak{Q}_0^C (\varphi_3) \mathfrak{Q}_0^D (\varphi_4) \right>_c \\ &= \Bigg[ \varepsilon^{ABE} \varepsilon^{CDE} \left( \mathrm{sgn} \left( \varphi_{12} \varphi_{24} \varphi_{43} \varphi_{31} \right) - \mathrm{sgn} \left( \varphi_{12} \varphi_{23} \varphi_{34} \varphi_{41} \right) \right) \\ &+ \varepsilon^{ACE} \varepsilon^{BDE} \left( \mathrm{sgn} \left( \varphi_{12} \varphi_{24} \varphi_{43} \varphi_{31} \right) - \mathrm{sgn} \left( \varphi_{13} \varphi_{32} \varphi_{24} \varphi_{41} \right) \right)  \\ &+ \varepsilon^{ADE} \varepsilon^{CBE} \left( \mathrm{sgn} \left( \varphi_{13} \varphi_{32} \varphi_{24} \varphi_{41} \right) - \mathrm{sgn} \left( \varphi_{12} \varphi_{23} \varphi_{34} \varphi_{41} \right) \right) \Bigg] \left( \frac{\Delta}{2} \right)^4 \frac{\rho (0) \sqrt{N}}{3 \pi}.
\end{split}
\end{align}
These expressions are analogous to correlation functions of the 1d vector topological QM, given by eq. (3.23) of \cite{MezeiPufuWang}, which has the known holographic $\mathrm{AdS}_2$ dual with $d_3^{ABC} = 0$ and $d_4^{ABCD} = 0$, described in Section 5 of \cite{MezeiPufuWang}. It is natural to expect that this 2d theory is a subsector of the bulk theory we consider.

\section{Bulk symmetry and dynamics}

One can extract the data needed to identify the symmetry algebra of the bulk theory from the correlation functions of the boundary topological theory.

The OPE of two currents in 1d is in general of the form \cite{MezeiPufuWang}
\begin{equation}
j^K (0) j^L(\varphi) = B \delta^{KL} + \frac{\mathrm{sgn} \varphi}{2} f^{KLM} j^M(0) + d^{KLM} j^M(0),
\end{equation}
where $B$ is a normalization constant, the indices $K$, $L$ and $M$ run over all conserved currents in the theory, $f^{KLM}$ is totally anti-symmetric, and $d^{KLM}$ is totally symmetric. These constants can be read from the correlators of currents as follows:
\begin{align}
\begin{split}
G_A &\equiv \left< j^{ [K} \left( \varphi_1 \right) j^{L]} \left( \varphi_2 \right) j^M \left( \varphi_3 \right) \right> = \frac{B}{2} f^{M [KL]}, \\
G_S &\equiv \left< j^{ (K} \left( \varphi_1 \right) j^{L)} \left( \varphi_2 \right) j^M \left( \varphi_3 \right) \right> = B d^{M (KL)}.
\end{split}
\end{align}

The holographic correspondence identifies the single-trace currents on the boundary with the bulk gauge fields, so in order to interpret the aforementioned 3-tensors as the gauge group structure constants of the 2d bulk non-Abelian gauge theory, and the symmetric tensor appearing in the action (\ref{eq:BulkAction}), the single-trace currents must form a closed algebra, without necessity to include any higher-trace currents.

\subsection{Single-trace operators}

The conserved single-trace currents $\mathfrak{X}_{\ell}$ we worked with, can be expanded in terms of $\mathfrak{X}_{\ell m}$ as
\begin{align}
\begin{split}
\mathfrak{X}_{\ell} \left( \varphi, y \right) = &\mathrm{i}^{\ell} \left( y^1 \right)^{\ell} \left( y^2 \right)^{\ell} \mathfrak{X}_{\ell 0} \left( \varphi \right) \\ &+\sum\limits_{ m = 1}^{\ell} \frac{\mathrm{i}^{\ell - m}}{\sqrt{2}} \left[ \left( y^1 \right)^{\ell + m} \left( y^2 \right)^{\ell - m} + \left( y^1 \right)^{\ell - m} \left( y^2 \right)^{\ell + m} \right] \mathfrak{X}_{\ell m} \left( \varphi \right) \\ &+ \sum\limits_{ m =  - \ell}^{-1} \frac{\mathrm{i}^{\ell - m - 1}}{\sqrt{2}} \left[ \left( y^1 \right)^{\ell + m} \left( y^2 \right)^{\ell - m} - \left( y^1 \right)^{\ell - m} \left( y^2 \right)^{\ell + m} \right] \mathfrak{X}_{\ell m} \left( \varphi \right).
\end{split}
\end{align}
From this expression it follows that
\begin{equation}
\mathfrak{X}_{\ell m} = \mathcal{D}_{\ell m} \mathfrak{X}_{\ell},
\end{equation}
where
\begin{align}
\begin{split}
\mathcal{D}_{\ell m} \left( y \right)=\begin{cases}
\frac{\left( - \mathrm{ i } \right)^{\ell - m}}{\sqrt{2} (\ell+ m)! (\ell - m)!} \left[ \frac{\partial^{2 \ell}}{\left( \partial y^1 \right)^{\ell + m} \left( \partial y^2 \right)^{\ell - m}} + \frac{\partial^{2 \ell}}{\left( \partial y^1 \right)^{\ell - m} \left( \partial y^2 \right)^{\ell + m}} \right], \qquad &m > 0, \\
\frac{\left( - \mathrm{ i } \right)^{\ell}}{\left( \ell! \right)^2} \frac{\partial^{2 \ell}}{\left( \partial y^1 \right)^{\ell} \left( \partial y^2 \right)^{\ell}},  &m=0, \\ \frac{\left( - \mathrm{ i } \right)^{\ell - m - 1}}{\sqrt{2} (\ell+ m)! (\ell - m)!} \left[ \frac{\partial^{2 \ell}}{\left( \partial y^1 \right)^{\ell + m} \left( \partial y^2 \right)^{\ell - m}} - \frac{\partial^{2 \ell}}{\left( \partial y^1 \right)^{\ell - m} \left( \partial y^2 \right)^{\ell + m}} \right], \qquad &m < 0.
\end{cases}
\end{split}
\end{align}

The 2- and 3-point functions of the operators defined this way are
\begin{align}
\begin{split}
\left< \mathfrak{X}_{\ell_1 m_1} \left( \varphi_1 \right) \mathfrak{X}_{\ell_2 m_2} \left( \varphi_2 \right) \right> = &B_{\ell_1 m_1}^{\mathfrak{XX}} \delta_{\ell_1 \ell_2} \delta_{m_1 m_2}, \\ \left< \mathfrak{X}_{\ell_1 m_1} \left( \varphi_1 \right) \mathfrak{X}_{\ell_2 m_2} \left( \varphi_2 \right) \mathfrak{X}_{\ell_3 m_3} \left( \varphi_3 \right) \right> = &C_{\ell_1 m_1, \ell_2 m_2, \ell_3 m_3}^{\mathfrak{XXX}} \\ &\times \left( \mathrm{sgn} \varphi_{21} \right)^{L_{12,3}} \left( \mathrm{sgn} \varphi_{31} \right)^{L_{13,2}} \left( \mathrm{sgn} \varphi_{32} \right)^{L_{23,1}},
\end{split}
\end{align}
where
\begin{align}
\begin{split}
B_{\ell m}^{\mathfrak{XX}} = \ &B_{\ell}^{\mathfrak{XX}} \mathcal{D}_{\ell m} \left( y_1 \right) \mathcal{D}_{\ell, -m} \left( y_2 \right) \left[\left< y_1 , y_2 \right>^{2 \ell} \right] = B_{\ell}^{\mathfrak{XX}} \frac{(2 \ell)!}{(\ell + m)! (\ell - m)!}, \\ C_{\ell_1 m_1, \ell_2 m_2, \ell_3 m_3}^{\mathfrak{XXX}} = \ &C_{\ell_1 \ell_2 \ell_3}^{\mathfrak{XXX}} \mathcal{D}_{\ell_1 m_1} \left( y_1 \right) \mathcal{D}_{\ell_2 m_2} \left( y_2 \right) \mathcal{D}_{\ell_3 m_3} \left( y_3 \right) \\ &\left[ \left< y_1, y_2 \right>^{\ell_1 + \ell_2 - \ell_3} \left< y_1, y_3 \right>^{\ell_1 + \ell_3 - \ell_2} \left< y_2, y_3 \right>^{\ell_2 + \ell_3 - \ell_1} \right].
\end{split}
\end{align}

After dividing $C_{\ell_1 m_1, \ell_2 m_2, \ell_3 m_3}^{\mathfrak{XXX}}$ by the normalization factor $\sqrt{B_{\ell_1 m_1}^{\mathfrak{XX}} B_{\ell_2 m_2}^{\mathfrak{XX}} B_{\ell_3 m_3}^{\mathfrak{XX}}}$, one obtains the anti-symmetric structure constants $f_{\ell_1 m_1, \ell_2 m_2, \ell_3 m_3}^{\mathfrak{XXX}}$, if $\ell_1 + \ell_2 + \ell_3$ is odd, and the symmetric 3-tensor $d_{\ell_1 m_1, \ell_2 m_2, \ell_3 m_3}^{\mathfrak{XXX}}$, related to the invariant symmetric 3-tensor of the group in a way we will uncover momentarily,\footnote{Mod factor of 2.} if $\ell_1 + \ell_2 + \ell_3$ is even. The algebra with the structure constants $f_{\ell_1 m_1, \ell_2 m_2, \ell_3 m_3}^{\mathfrak{XXX}}$ can be shown \cite{PopeRomansShen,JoungMkrtchyan} to be the algebra of area-preserving diffeomorphisms of the 2-sphere.

In our $k \neq 1$ theory, we must add the single-trace currents $\mathfrak{Q}_{\ell m}^A$ to the algebra. The 2- and 3-point functions we must consider are
\begin{align}
\begin{split}
\left< \mathfrak{Q}^{A_1}_{\ell_1 m_1} \left( \varphi_1 \right) \mathfrak{Q}^{A_2}_{\ell_2 m_2} \left( \varphi_2 \right) \right> = &\mathrm{Tr} \left( T^{A_1} T^{A_2} \right) B_{\ell_1 m_1}^{\mathfrak{QQ}} \delta_{\ell_1 \ell_2} \delta_{m_1 m_2}, \\ \left< \mathfrak{Q}^{A_1}_{\ell_1 m_1} \left( \varphi_1 \right) \mathfrak{Q}^{A_2}_{\ell_2 m_2} \left( \varphi_2 \right) \mathfrak{Q}^{A_3}_{\ell_3 m_3} \left( \varphi_3 \right) \right> = &\mathrm{i} \mathrm{Tr} \left( T^{A_1} \left[ T^{A_2} , T^{A_3} \right] \right) C_{\ell_1 m_1, \ell_2 m_2, \ell_3 m_3}^{\mathfrak{QQQ}} \\ &\times \left( \mathrm{sgn} \varphi_{21} \right)^{L_{12,3}+1} \left( \mathrm{sgn} \varphi_{31} \right)^{L_{13,2}+1} \left( \mathrm{sgn} \varphi_{32} \right)^{L_{23,1}+1}, \\ \left< \mathfrak{Q}^{A_1}_{\ell_1 m_1} \left( \varphi_1 \right) \mathfrak{Q}^{A_2}_{\ell_2 m_2} \left( \varphi_2 \right) \mathfrak{X}_{\ell_3 m_3} \left( \varphi_3 \right) \right> = &\mathrm{Tr} \left( T^{A_1} T^{A_2} \right) C_{\ell_1 m_1, \ell_2 m_2, \ell_3 m_3}^{\mathfrak{QQX}} \\ &\times \left( \mathrm{sgn} \varphi_{21} \right)^{L_{12,3}+2} \left( \mathrm{sgn} \varphi_{31} \right)^{L_{13,2}} \left( \mathrm{sgn} \varphi_{32} \right)^{L_{23,1}},
\end{split}
\end{align}
where $B_{\ell_1 m_1}^{\mathfrak{QQ}}$ and $C_{\ell_1 m_1, \ell_2 m_2, \ell_3 m_3}^{\mathfrak{QQQ}}$ are defined similarly to $B_{\ell_1 m_1}^{\mathfrak{XX}}$ and $C_{\ell_1 m_1, \ell_2 m_2, \ell_3 m_3}^{\mathfrak{XXX}}$, and are given by
\begin{align}
\begin{split}
B_{\ell m}^{\mathfrak{QQ}} = \ &B_{\ell}^{\mathfrak{QQ}} \mathcal{D}_{\ell m} \left( y_1 \right) \mathcal{D}_{\ell, -m} \left( y_2 \right) \left[\left< y_1 , y_2 \right>^{2 \ell} \right] = B_{\ell}^{\mathfrak{QQ}} \frac{(2 \ell)!}{(\ell + m)! (\ell - m)!}, \\ C_{\ell_1 m_1, \ell_2 m_2, \ell_3 m_3}^{\mathfrak{QQQ}} = \ &C_{\ell_1 \ell_2 \ell_3}^{\mathfrak{QQQ}} \mathcal{D}_{\ell_1 m_1} \left( y_1 \right) \mathcal{D}_{\ell_2 m_2} \left( y_2 \right) \mathcal{D}_{\ell_3 m_3} \left( y_3 \right) \\ &\left[ \left< y_1, y_2 \right>^{\ell_1 + \ell_2 - \ell_3} \left< y_1, y_3 \right>^{\ell_1 + \ell_3 - \ell_2} \left< y_2, y_3 \right>^{\ell_2 + \ell_3 - \ell_1} \right].
\end{split}
\end{align}

Again, one can extract the structure constants $f_{\ell_1 m_1 A_1,\ell_2 m_2 A_2,\ell_3 m_3 A_3}^{\mathfrak{QQQ}}$, and the symmetric 3-tensor $d_{\ell_1 m_1 A_1,\ell_2 m_2 A_2,\ell_3 m_3 A_3}^{\mathfrak{QQQ}}$ from these correlation functions. The result is
\begin{align} \label{eq:QQQStructureConstants}
\begin{split}
f_{\ell_1 m_1 A_1,\ell_2 m_2 A_2,\ell_3 m_3 A_3}^{\mathfrak{QQQ}} = &\frac{f_{A_1 A_2 A_3} C_{\ell_1 \ell_2 \ell_3}^{\mathfrak{QQQ}}}{\sqrt{B_{\ell_1 m_1}^{\mathfrak{QQ}} B_{\ell_2 m_2}^{\mathfrak{QQ}} B_{\ell_3 m_3}^{\mathfrak{QQ}}}} \mathcal{D}_{\ell_1 m_1} \left( y_1 \right) \mathcal{D}_{\ell_2 m_2} \left( y_2 \right) \mathcal{D}_{\ell_3 m_3} \left( y_3 \right) \\ &\left[ \left< y_1, y_2 \right>^{\ell_1 + \ell_2 - \ell_3} \left< y_1, y_3 \right>^{\ell_1 + \ell_3 - \ell_2} \left< y_2, y_3 \right>^{\ell_2 + \ell_3 - \ell_1} \right],
\end{split}
\end{align}
if $\ell_1 + \ell_2 + \ell_3$ is even. The symmetric 3-tensor is given by the same expression, but for the odd $\ell_1 + \ell_2 + \ell_3$. Here $f_{A_1 A_2 A_3}$ are the structure constants of the algebra $\mathfrak{su} \left( k \right)$, and are given as usual by $f_{A_1 A_2 A_3} = - \mathrm{i} \mathrm{Tr} \left( T^{A_1} \left[ T^{A_2} , T^{A_3} \right] \right)$.

The discussion of $f_{\ell_1 m_1 A_1,\ell_2 m_2 A_2,\ell_3 m_3}^{\mathfrak{QQX}}$ and $d_{\ell_1 m_1 A_1,\ell_2 m_2 A_2,\ell_3 m_3}^{\mathfrak{QQX}}$ is completely analogous. The structure constants are
\begin{align} \label{eq:QQXStructureConstants}
\begin{split}
f_{\ell_1 m_1 A_1,\ell_2 m_2 A_2,\ell_3 m_3}^{\mathfrak{QQX}} = &\frac{\delta_{A_1 A_2} C_{\ell_1 \ell_2 \ell_3}^{\mathfrak{QQX}}}{\sqrt{B_{\ell_1 m_1}^{\mathfrak{QQ}} B_{\ell_2 m_2}^{\mathfrak{QQ}} B_{\ell_3 m_3}^{\mathfrak{XX}}}} \mathcal{D}_{\ell_1 m_1} \left( y_1 \right) \mathcal{D}_{\ell_2 m_2} \left( y_2 \right) \mathcal{D}_{\ell_3 m_3} \left( y_3 \right) \\ &\left[ \left< y_1, y_2 \right>^{\ell_1 + \ell_2 - \ell_3} \left< y_1, y_3 \right>^{\ell_1 + \ell_3 - \ell_2} \left< y_2, y_3 \right>^{\ell_2 + \ell_3 - \ell_1} \right],
\end{split}
\end{align}
if $\ell_1 + \ell_2 + \ell_3$ is odd. The symmetric 3-tensor $d_{\ell_1 m_1 A_1,\ell_2 m_2 A_2,\ell_3 m_3}^{\mathfrak{QQX}}$ is given by the same expression, but for the even $\ell_1 + \ell_2 + \ell_3$.

\subsection{Jacobi identities}

As we already mentioned, in order to map the single-trace currents in the boundary theory to the gauge fields in the bulk, the currents must form a closed algebra. It means that at the leading order in $N$, the structure constants we obtain must satisfy the Jacobi identities, and the symmetric 3-tensors must obey the invariance condition
\begin{equation} \label{eq:SingleTraceInvariance}
f^{\mathfrak{XXX}}_{x_1 x_5 x_2} d^{\mathfrak{XXX}}_{x_3 x_4 x_5} + f^{\mathfrak{XXX}}_{x_1 x_5 x_3} d^{\mathfrak{XXX}}_{x_4 x_2 x_5} + f^{\mathfrak{XXX}}_{x_1 x_5 x_4} d^{\mathfrak{XXX}}_{x_2 x_3 x_5} = 0,
\end{equation}
and its generalizations involving $\mathfrak{Q}$'s, in order to get a gauge-invariant cubic interaction in (\ref{eq:BulkAction}). Here $x_i = \left( \ell_i , m_i \right)$ .

From now on, we work with the normalization of the single-trace operators in which the 2-point functions are normalized to $\delta_{KL}$. Working with the normalized structure constants allows us not to distinguish the upper and the lower indices, and in this section we will use only the lower ones. In this normalization, the 3-point functions scale as $\left< \mathfrak{X} \mathfrak{X} \mathfrak{X} \right> \sim N^{-\frac{3}{4}}$, $\left< \mathfrak{Q} \mathfrak{Q} \mathfrak{X} \right> \sim N^{-\frac{3}{4}}$ and $\left< \mathfrak{Q} \mathfrak{Q} \mathfrak{Q} \right> \sim N^{-\frac{1}{4}}$, and at the leading order in $N$, the Jacobi identities for the currents $\mathfrak{X}_{\ell m}$ and $\mathfrak{Q}_{\ell m}^A$ give rise to the following relations for the structure constants:
\begin{align}
\begin{split}
f^{\mathfrak{XXX}}_{x_1 x_2 x_5} f^{\mathfrak{XXX}}_{x_3 x_5 x_4} + f^{\mathfrak{XXX}}_{x_2 x_3 x_5} f^{\mathfrak{XXX}}_{x_1 x_5 x_4} + f^{\mathfrak{XXX}}_{x_3 x_1 x_5} f^{\mathfrak{XXX}}_{x_2 x_5 x_4} &= 0, \\
f^{\mathfrak{QQQ}}_{q_1 q_2 q_5} f^{\mathfrak{QQQ}}_{q_3 q_5 q_4} + f^{\mathfrak{QQQ}}_{q_2 q_3 q_5} f^{\mathfrak{QQQ}}_{q_1 q_5 q_4} + f^{\mathfrak{QQQ}}_{q_3 q_1 q_5} f^{\mathfrak{QQQ}}_{q_2 q_5 q_4} &= 0, \\
f^{\mathfrak{QQQ}}_{q_1 q_2 q_5} f^{\mathfrak{QQX}}_{q_3 q_5 x_4} + f^{\mathfrak{QQQ}}_{q_2 q_3 q_5} f^{\mathfrak{QQX}}_{q_1 q_5 x_4} + f^{\mathfrak{QQQ}}_{q_3 q_1 q_5} f^{\mathfrak{QQX}}_{q_2 q_5 x_4} &= 0, \\
f^{\mathfrak{QQX}}_{q_2 q_3 x_5} f^{\mathfrak{XXX}}_{x_1 x_5 x_4} - f^{\mathfrak{QQX}}_{q_2 q_5 x_1} f^{\mathfrak{QQX}}_{q_3 q_5 x_4} - f^{\mathfrak{QQX}}_{q_3 q_5 x_1} f^{\mathfrak{QQX}}_{q_2 q_5 x_4} &= 0,
\end{split}
\end{align}
where $q_j = \left( \ell_j , m_j , A_j \right)$. All these identities are satisfied with our 3-point functions at the leading order in $N$. But the invariance condition (\ref{eq:SingleTraceInvariance}) is not satisfied even at the leading order. The origin of this is the following. If we consider the $d_3$-invariance condition of the full boundary current algebra, and take all 4 external indices to be single-trace, for the purely $\mathfrak{X}$ curents we get\footnote{The generalization for $\mathfrak{Q}$'s is obvious.}
\begin{equation} \label{eq:d3Invariance}
g^{KL} f^{\mathfrak{XXX}}_{x_1 K x_2} d^{\mathfrak{XXX}}_{x_3 x_4 L} + g^{KL} f^{\mathfrak{XXX}}_{x_1 Kx_3} d^{\mathfrak{XXX}}_{x_4 x_2 L} + g^{KL} f^{\mathfrak{XXX}}_{x_1 K x_4} d^{\mathfrak{XXX}}_{x_2 x_3 L} = 0,
\end{equation}
where again the indices $K$ and $L$ run over all currents in the theory (not only single-trace). In contrast to the identities for the structure constants, these equalities are not closed on the single-trace sector, which is not surprising, because the mixed single- and the double-trace contribution to the identity has the same $\mathcal{O} \left( N^{- \frac{3}{2}} \right)$ order, if $d^{\mathfrak{XXX}}_{x_2 x_3 M}$ is reduced to product of two 2-point functions, as we show in the next subsection. Let's analyze this doube-trace contribution.

\subsection{The double-trace contribution}

We introduce the double-trace operators $\mathfrak{X}_{ \left[ x_1 x_2 \right]} \left( \varphi \right) = : \mathfrak{X}_{x_1} \left( \varphi \right) \mathfrak{X}_{x_2} \left( \varphi \right) :$, where the normal ordering is defined as
\begin{align}
\begin{split}
: \mathfrak{X}_{x_1} \left( \varphi \right) \mathfrak{X}_{x_2} \left( \varphi \right) : = &\frac{ \mathfrak{X}_{x_1} \left( \varphi \right) \mathfrak{X}_{x_2} \left( \varphi + \varepsilon \right) + \mathfrak{X}_{x_1} \left( \varphi \right) \mathfrak{X}_{x_2} \left( \varphi - \varepsilon \right) }{2} \\ &- \left< \frac{ \mathfrak{X}_{x_1} \left( \varphi \right) \mathfrak{X}_{x_2} \left( \varphi + \varepsilon \right) + \mathfrak{X}_{x_1} \left( \varphi \right) \mathfrak{X}_{x_2} \left( \varphi - \varepsilon \right) }{2} \right>.
\end{split}
\end{align}
Here $\varepsilon$ is set to zero at the end of any computation, and the complete set of independent operators is defined by $\ell_2 \geq \ell_1$. The double-trace operators $\mathfrak{Q}_{ \left[ q_1 q_2 \right] } \left( \varphi \right)= : \mathfrak{Q}_{q_1} \left( \varphi \right) \mathfrak{Q}_{q_2} \left( \varphi \right) :$, and $\left( \mathfrak{Q} \mathfrak{X} \right)_{ \left[ q_1 x_2 \right] } \left( \varphi \right) = : \mathfrak{Q}_{q_1} \left( \varphi \right) \mathfrak{X}_{x_2} \left( \varphi \right) :$ are defined analogously.

The double-trace operators are normalized as
\begin{equation}
\left< \mathfrak{X}_{\left[ x_1  x_2 \right]} \left( \varphi \right) \mathfrak{X}_{\left[ x_1  x_2 \right]} \left( \varphi \right) \right> = B_{x_1}^{\mathfrak{XX}} B_{x_2}^{\mathfrak{XX}} \left( 1 + \delta_{x_1 x_2} \right),
\end{equation}
their connected 3-point functions with 2 single-trace operators scale as $\left< \mathfrak{X}_{x_1} \mathfrak{X}_{x_2} \mathfrak{X}_{ \left[ x_3 x_4 \right] } \right>_c \sim N^{-\frac{3}{2}}$, and $\left< \mathfrak{X}_{x_1} \mathfrak{X}_{x_2} \mathfrak{X}_{\left[ x_1 x_2 \right]} \right> \sim N^{0}$. From the latter relation it follows that the mixed single-trace/double-trace contribution to (\ref{eq:d3Invariance}) is of the same $\mathcal{O} \left( N^{-\frac{3}{2}} \right)$ order as the purely single-trace contribution.

In order to make the ``single-trace'' algebra closed at the leading order in $N$,\footnote{We expect that it is possible to get the closed algebra at any order in $N$ by mixing the single-trace currents with appropriate combination of higher-trace currents.} we re-define what we mean by the single-trace operators as\footnote{A necessity of doing this is mentioned in \cite{MezeiPufuWang}.}
\begin{equation} \label{eq:NewSingleTracesX}
\overline{\mathfrak{X}}_{x_1} = \mathfrak{X}_{x_1} + \frac{1}{N^{\frac{3}{4}}} a_{x_1 \left[ x_2  x_3 \right]} \mathfrak{X}_{\left[x_2  x_3 \right]}.
\end{equation}
Here, the coefficients $a_{x_1 \left[ x_5 x_6 \right]}$ don't depend on $N$. In this new basis, the 3-point function of our new single-trace operators having the $\mathcal{O} \left( N^{-\frac{3}{4}} \right)$ order, changes as
\begin{align} \label{eq:New3PointFunction}
\begin{split}
&\left< \overline{\mathfrak{X}}_{x_1} \overline{\mathfrak{X}}_{x_2} \overline{\mathfrak{X}}_{x_3} \right> = \left< \mathfrak{X}_{x_1} \mathfrak{X}_{x_2} \mathfrak{X}_{x_3} \right> \\ &+ \frac{1}{N^{\frac{3}{4}}} \Bigg[ a_{x_1 \left[x_4 x_5 \right]} \left< \mathfrak{X}_{\left[ x_4 x_5 \right]} \mathfrak{X}_{x_2} \mathfrak{X}_{x_3} \right> + a_{x_2 \left[ x_4 x_5 \right]} \left< \mathfrak{X}_{x_1} \mathfrak{X}_{\left[x_4 x_5 \right]} \mathfrak{X}_{x_3} \right> + a_{x_3 \left[ x_4 x_5 \right]} \left< \mathfrak{X}_{x_1} \mathfrak{X}_{x_2} \mathfrak{X}_{\left[ x_4 x_5 \right]} \right> \Bigg].
\end{split}
\end{align}
The term in the square brackets contribute at the order we work with only when the correlators reduce to products of the 2-point functions, leading to an expression independent of coordinates, which means that only the symmetric tensors $d^{\mathfrak{XXX}}_{x_1 x_2 x_3}$ are affected by our change of operators, while the structure constants stay the same. In order to get the algebra closed on the single-trace operators, we want our $\overline{d^{\mathfrak{XXX}}_{x_1 x_2 x_3}}$ to be the genuine invariant symmetric 3-tensor of the group, defined as $ \overline{d^{\mathfrak{XXX}}_{x_1 x_2 x_3}} = \mathrm{Tr} \left( T_{x_1} \left\{ T_{x_2} , T_{x_3} \right\} \right)$,\footnote{We didn't manage to compute these constants in the closed form.} where $\left( T_{x_k} \right)_{x_m x_n} = - \mathrm{i} f^{\mathfrak{XXX}}_{x_k x_m x_n}$. By choosing the coefficients $a_{x_1 \left[x_2 x_3 \right]}$ to be totally symmetric in all indices, we get from (\ref{eq:New3PointFunction})
\begin{equation} \label{eq:ExplicitA}
\overline{d^{\mathfrak{XXX}}_{x_1 x_2 x_3}} = d^{\mathfrak{XXX}}_{x_1 x_2 x_3} + 3 \frac{a_{x_1 x_2 x_3}}{N^{\frac{3}{4}}},
\end{equation}
from which it follows that if\footnote{Equation (\ref{eq:ExplicitA}) is valid when all $x$'s are different. When two of them are equal, the coefficient of the second term is 4 instead of 3, and if all of them are equal, the coefficient is 6.}
\begin{equation}
a_{x_1 x_2 x_3} = \frac{N^{\frac{3}{4}}}{3} \left[ \overline{d^{\mathfrak{XXX}}_{x_1 x_2 x_3}} - d^{\mathfrak{XXX}}_{x_1 x_2 x_3} \right],
\end{equation}
the algebra of single trace-operators is closed.

The situation with operators involving $\mathfrak{Q}$'s is similar. We re-define the single-trace operators as
\begin{equation} \label{eq:NewSingleTracesQ}
\overline{\mathfrak{Q}}_{q_1} = \mathfrak{Q}_{q_1} + \frac{1}{N^{\frac{1}{4}}} b_{q_1 \left[ q_2 q_3 \right]} \mathfrak{Q}_{ \left[ q_2 q_3 \right] } + \frac{1}{N^{\frac{3}{4}}} c_{q_1 \left[ q_2 x_2 \right] } \left( \mathfrak{Q} \mathfrak{X} \right)_{ \left[ q_2 x_2 \right] }.
\end{equation}

The 3-point function of the single-trace operators is now
\begin{align}
\begin{split}
&\left< \overline{\mathfrak{Q}}_{q_1} \overline{\mathfrak{Q}}_{q_2} \overline{\mathfrak{Q}}_{q_3} \right> = \left< \mathfrak{Q}_{q_1} \mathfrak{Q}_{q_2} \mathfrak{Q}_{q_3} \right> \\ &+ \frac{1}{N^{\frac{3}{4}}} \Bigg[ c_{q_1 \left[q_4 x_4 \right]} \left< \left( \mathfrak{Q} \mathfrak{X} \right)_{\left[ q_4 x_4 \right]} \mathfrak{Q}_{q_2} \mathfrak{Q}_{q_3} \right> + c_{q_2 \left[ q_4 x_4 \right]} \left< \mathfrak{Q}_{q_1} \left( \mathfrak{Q} \mathfrak{X} \right)_{\left[ q_4 x_4 \right]} \mathfrak{Q}_{q_3} \right> + c_{q_3 \left[q_4 x_4 \right]} \left< \mathfrak{Q}_{q_1} \mathfrak{Q}_{q_2} \left( \mathfrak{Q} \mathfrak{X} \right)_{\left[ q_4 x_4 \right]} \right> \Bigg] \\ &+ \frac{1}{N^{\frac{1}{4}}} \Bigg[ b_{q_1 \left[q_4 q_5 \right]} \left< \mathfrak{Q}_{\left[ q_4 q_5 \right]} \mathfrak{Q}_{q_2} \mathfrak{Q}_{q_3} \right> + b_{q_2 \left[ q_4 q_5 \right]} \left< \mathfrak{Q}_{q_1} \mathfrak{Q}_{\left[q_4 q_5 \right]} \mathfrak{Q}_{q_3} \right> + b_{q_3 \left[ q_4 q_5 \right]} \left< \mathfrak{Q}_{q_1} \mathfrak{Q}_{q_2} \mathfrak{Q}_{\left[ q_4 q_5 \right]} \right> \Bigg],
\end{split}
\end{align}
and
\begin{align}
\begin{split}
&\left< \overline{\mathfrak{Q}}_{q_1} \overline{\mathfrak{Q}}_{q_2} \overline{\mathfrak{X}}_{x_3} \right> = \left< \mathfrak{Q}_{q_1} \mathfrak{Q}_{q_2} \mathfrak{X}_{x_3} \right> \\ &+ \frac{1}{N^{\frac{3}{4}}} \Bigg[ c_{q_1 \left[q_4 x_4 \right]} \left< \left( \mathfrak{Q} \mathfrak{X} \right)_{\left[ q_4 x_4 \right]} \mathfrak{Q}_{q_2} \mathfrak{X}_{x_3} \right> + c_{q_2 \left[ q_4 x_4 \right]} \left< \mathfrak{Q}_{q_1} \left( \mathfrak{Q} \mathfrak{X} \right)_{\left[ q_4 x_4 \right]} \mathfrak{X}_{x_3} \right> + a_{x_3 \left[x_4 x_5 \right]} \left< \mathfrak{Q}_{q_1} \mathfrak{Q}_{q_2} \mathfrak{X}_{\left[ x_4 x_5 \right]} \right> \Bigg] \\ &+ \frac{1}{N^{\frac{1}{4}}} \Bigg[ b_{q_1 \left[q_4 q_5 \right]} \left< \mathfrak{Q}_{\left[ q_4 q_5 \right]} \mathfrak{Q}_{q_2} \mathfrak{X}_{x_3} \right> + b_{q_2 \left[ q_4 q_5 \right]} \left< \mathfrak{Q}_{q_1} \mathfrak{Q}_{\left[q_4 q_5 \right]} \mathfrak{X}_{x_3} \right> \Bigg],
\end{split}
\end{align}
Taking into account that the normalized correlators $\left< \mathfrak{Q} \mathfrak{Q} \mathfrak{Q} \mathfrak{X} \right>$ and $\left< \mathfrak{Q} \mathfrak{Q} \mathfrak{X} \mathfrak{X} \right>$ scale as $N^{-1}$ and $N^{-\frac{3}{2}}$, respectively, and choosing $b$'s and $c$'s to be totally symmetric, we find that the algebra closes if\footnote{The story with the coefficient 3 in the first equation is the same as in the case of purely $\mathfrak{X}$ operators.}
\begin{equation}
b_{q_1 q_2 q_3 } = \frac{N^{\frac{1}{4}}}{3} \left[ \overline{d^{\mathfrak{QQQ}}_{q_1 q_2 q_3}} - d^{\mathfrak{QQQ}}_{q_1 q_2 q_3} \right], \quad c_{q_1 q_2 x_3} = \frac{N^{\frac{3}{4}}}{2} \left[ \overline{d^{\mathfrak{QQX}}_{q_1 q_2 x_3}} - d^{\mathfrak{QQX}}_{q_1 q_2 x_3} \right].
\end{equation}

Note that the invariant symmetric tensors $\overline{d^{\mathfrak{QQQ}}_{q_1 q_2 q_3}}$ are proportional to the symmetric 3-tensors $d_{ABC}$ of $\mathrm{SU} \left( k \right)$ in contrast to the naive symmetric tensors $d^{\mathfrak{QQQ}}_{q_1 q_2 q_3}$, which are proportional to $f_{ABC}$ (see (\ref{eq:3pointfunction})).

Our holographic proposal is that the modified single-trace operators (\ref{eq:NewSingleTracesX}) and (\ref{eq:NewSingleTracesQ}) must be mapped to the non-Abelian gauge fields in the bulk.

The fact that the action of the bulk theory defined on $\mathrm{AdS}_2$, contains the terms of two different types --- one of them has an $N^{\frac{3}{2}}$-dependence, and the other one is proportional to $\sqrt{N}$ --- can be understood from the 11d perspective. The 1d theory we considered, captures dynamics of the protected topological sector of 3d $\mathcal{N} = 4$ theory with $k$ hypermultiplets, which has M-theory in the $\mathrm{AdS}_4 \times \mathrm{S}^7/\mathbb{Z}_k$ background with $\mathrm{SU} (k)$ gauge fields living on the fixed points on the orbifold, as a gravity dual \cite{FerraraKehigiasPartoucheZaffaroni,Gomis}. The action of this theory in a given background is of the form $S \sim \frac{1}{\ell_{Pl}^9} \int \mathrm{d}^{11} x \sqrt{g} R + \frac{1}{\ell_{Pl}^3} \int \mathrm{d}^7 x \sqrt{g} \mathrm{Tr} \left( F_{\mu \nu}^A F^{\mu \nu A} \right) + \cdots$. If the bulk fields $\mathfrak{Q}_{\ell m}^A$ arise as the Kaluza-Klein modes of the 7d theory, the $\sqrt{N}$-behavior of the terms with $q_i$ indices in the 2d action, makes sense.

\section{Conclusions and future directions}

In this paper, we proposed a holographic dual to the flavored topological quantum mechanics with non-zero FI term. There are some open question left though.

There is a string theory construction of the 1d topological quantum mechanics we considered ---  one must place stacks of $N$ D2-branes and $k$ D6-branes in the $\Omega$-background in Type IIA string theory \cite{CostelloMTheoryOmegaBackground,HolographyAndKoszulDuality}. It would be great to understand a relation of this brane construction to our $\mathrm{AdS}_2$ dual in spirit of the D-brane near-horizon limit in the canonical examples of AdS/CFT correspondence.

We computed the structure constants of the bulk theory gauge group, but haven't identified the group. It is also very interesting to understand the geometrical meaning of the group.

We would also like to understand a relation of our construction to the program of topological holography \cite{HolographyAndKoszulDuality,IshtiaqueMoosavianZhou,CostelloGaiotto,GaiottoMTheory}.

\section*{Acknowledgements}

We would like to thank Yifan Wang for valuable discussions and comments on a draft of this manuscript, and especially Ofer Aharony for useful discussions, general guidance and comments on a draft of this manuscript. This work was supported in part by an Israel Science Foundation center for excellence grant (grant number 1989/14) and by the Minerva foundation with funding from the Federal German Ministry for Education and Research.


\newpage

\addcontentsline{toc}{section}{References}

\begin{thebibliography}{99}

\bibitem{MezeiPufuWang}
M.~Mezei, S.~S.~Pufu and Y.~Wang,
\newblock {``A 2d/1d Holographic Duality},''
\newblock [arXiv:1703.08749].

\bibitem{DedushenkoPufuYacoby}
M.~Dedushenko, S.~S.~Pufu and R.~Yacoby,
\newblock {``A One-dimensional Theory for Higgs Branch Operators},''
\newblock JHEP \textbf{1803} (2018) 138
\newblock [arXiv:1610.00740].

\bibitem{ChesterLeePufuYacoby}
S.~M.~Chester, J.~Lee, S.~S.~Pufu and R.~Yacoby,
\newblock {``Exact Correlators of BPS Operators from the 3d Superconformal Bootstrap},''
\newblock JHEP \textbf{1503} (2015) 130
\newblock [arXiv:1412.0334].

\bibitem{BeemPeelaersRastelli}
C.~Beem, W.~Peelaers and L.~Rastelli,
\newblock {``Deformation Quantization and Superconformal Symmetry in Three Dimensions},''
\newblock Commun. Math. Phys. \textbf{354} (2017) no.1, 345
\newblock [arXiv:1601.05378].

\bibitem{TwistedSupergravityAndItsQuantization}
K.~Costello and S.~Li,
\newblock {``Twisted Supergravity and its Quantization},''
\newblock [arXiv:1606.00365].

\bibitem{CostelloMTheoryOmegaBackground}
K.~Costello,
\newblock {``M-theory in the $\Omega$-background and 5-dimensional Non-commutative Gauge Theory},''
\newblock [arXiv:1610.04144].

\bibitem{HolographyAndKoszulDuality}
K.~Costello,
\newblock {``Holography and Koszul Duality: the Example of the M2 Brane},''
\newblock [arXiv:1705.02500].

\bibitem{GaiottoAbajian}
D.~Gaiotto and J.~Abajian,
\newblock {``Twisted M2 Brane Holography and Sphere Correlation Functions},''
\newblock [arXiv:2004.13810].

\bibitem{MaldacenaHol}
J.~Maldacena,
\newblock {``The Large N Limit of Superconformal Field Theories and Supergravity},''
\newblock Adv. Theor. Math. Phys. \textbf{2} (1998) 231-252
\newblock [arXiv:hep-th/9711200].

\bibitem{WittenHol}
E.~Witten,
\newblock {``Anti de Sitter Space and Holography},''
\newblock Adv. Theor. Math. Phys. \textbf{2} (1998) 253-291
\newblock [arXiv:hep-th/9802150].

\bibitem{GKPHol}
S.~Gubser, I.~Klebanov and A.~Polyakov,
\newblock {``Gauge Theory Correlators from Noncritical String Theory},''
\newblock Phys. Lett. \textbf{B428} (1998) 105-114
\newblock [arXiv:hep-th/9802109].

\bibitem{ChernSimonsReview}
M.~Mari\~{n}o,
\newblock {``Lectures on Localization and Matrix Models in Supersymmetric Chern-Simons-Matter Theories},''
\newblock J. Phys. A \textbf{44} (2011) 463001
\newblock [arXiv:1104.0783].

\bibitem{MezeiPufu}
M.~Mezei and S.~Pufu,
\newblock {``Three-sphere Free Energy for Classical Gauge Groups},''
\newblock JHEP \textbf{02} (2014) 037
\newblock [arXiv:1312.0920].

\bibitem{GrassiMarino}
A.~Grassi and M.~Mari\~{n}o,
\newblock {``M-theoretic Matrix Models},''
\newblock JHEP \textbf{02} (2015) 115
\newblock [arXiv:1403.4276].

\bibitem{ABJM}
O.~Aharony, O.~Bergman, D.L.~Jafferis and J.~Maldacena,
\newblock {``$\mathcal{N} = 6$ Superconformal Chern-Simons-Matter Theories, M2-branes and their Gravity Duals},''
\newblock JHEP \textbf{10} (2008) 091
\newblock [arXiv:0806.1218].

\bibitem{PopeRomansShen}
C.~N.~Pope, L.~J.~Romans and X.~Shen,
\newblock {``A New Higher Spin Algebra and the Lone Star Product},''
\newblock Phys. Lett. B \textbf{242} (1990) 401.

\bibitem{JoungMkrtchyan}
E.~Joung and K.~Mkrtchyan,
\newblock {``Notes on Higher-spin Algebras: Minimal Representations and Structure Constants},''
\newblock JHEP \textbf{1405} (2014) 103
\newblock [arXiv:1401.7977].

\bibitem{FerraraKehigiasPartoucheZaffaroni}
S.~Ferrara, A.~Kehagias, H.~Partouche and A.~Zaffaroni,
\newblock {``Membranes and Five-branes with Lower Supersymmetry and their AdS Supergravity Duals},''
\newblock Phys. Lett. B \textbf{431} (1998)
\newblock [arXiv:hep-th/9803109].

\bibitem{Gomis}
J.~Gomis,
\newblock {``Anti de Sitter Geometry and Strongly Coupled Gauge Theories},''
\newblock Phys. Lett. B \textbf{435} (1998) 299-302
\newblock [arXiv:hep-th/9803119 ].

\bibitem{IshtiaqueMoosavianZhou}
N.~Ishtiaque, S.~Faroogh~Moosavian and Y.~Zhou,
\newblock {``Topological Holography: the Example of the D2-D4 Brane System},''
\newblock [arXiv:1809.00372].

\bibitem{CostelloGaiotto}
K.~Costello and D.~Gaiotto,
\newblock {``Twisted Holography},''
\newblock [arXiv:1812.09257].

\bibitem{GaiottoMTheory}
D.~Gaiotto and J.~Oh,
\newblock {``Aspects of $\Omega$-deformed M-theory},''
\newblock [arXiv:1907.06495].

\end{thebibliography}
\end{document}